\documentclass{ws-mpla}

\newcommand{\eq}[1]{(\ref{#1})}
\newcommand{\beqn}{\begin{eqnarray}}
\newcommand{\eeqn}{\end{eqnarray}}
\def\bbbone{{\mathchoice {\rm 1\mskip-4mu l} {\rm 1\mskip-4mu l}{\rm 1\mskip-4.5mu l} {\rm 1\mskip-5mu l}}}
\newcommand{\cL}{{\cal L}}
\newcommand{\cD}{{\cal D}}
\newcommand{\cF}{{\cal F}}
\newcommand{\cA}{{\cal A}}
\newcommand{\cC}{{\cal C}}

\begin{document}

\markboth{M. N. Chernodub}{Monopoles from Quark Condensates in QCD}

\catchline{}{}{}{}{}

\title{
\vskip -5mm
MONOPOLES FROM QUARK CONDENSATES IN QCD\thanks{Talks given
at International Conference on Chiral Symmetry in Hadron and Nuclear Physics (Chiral07)
November 13-16, 2007, Osaka University, Japan and at the session of Russian
Academy of Sciences "Physics of Fundamental Interactions",
ITEP, Moscow, November 26-30, 2007.}\\
\vskip -30mm
\rightline{\includegraphics[width=2.5cm]{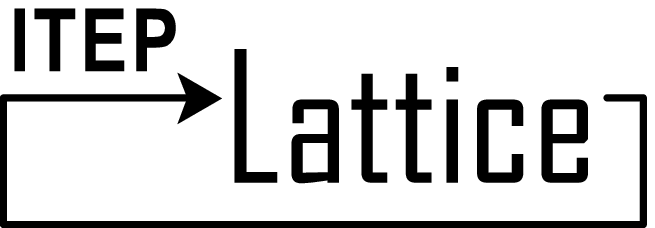}}
\vskip 3mm
\rightline{\large\textmd{ITEP-LAT/2008-06}}
\vskip 12mm
}

\author{\footnotesize M. N. Chernodub}

\address{Institute for Theoretical and Experimental Physics,\\
B.Cheremushkinskaya 25, Moscow, Russia\\
Maxim.Chernodub@itep.ru}

\maketitle

\pub{Received (Day Month Year)}{Revised (Day Month Year)}

\begin{abstract}
Chiral monopoles are hedgehoglike structures in local chiral
condensates in QCD. These monopoles are (i) made of quark and gluon
fields; (ii) explicitly gauge-invariant; and (iii) they carry
quantized and conserved chromomagnetic charge. We argue that the
chiral condensate vanishes in a core of the chiral monopole while the
density of these monopoles increases with temperature wiping out the
quark condensate in quark-gluon plasma. We suggest that the dynamics of the
chiral monopoles is responsible for the chiral symmetry restoration in high
temperature phase of QCD. We also argue that the chiral monopoles are unlikely
to be associated with confining degrees of freedom. In our approach the chiral symmetry
restoration and the color deconfinement in QCD are not necessarily related
to each other and the corresponding transitions may occur at different temperatures.
\end{abstract}

\ccode{25.75.Nq, 11.30.Rd, 14.80.Hv}

\section{Introduction}

The phase structure of Quantum Chromodynamics (QCD) at non-zero temperature and
finite chemical potential attracts increasing attention these
days.\cite{ref:plasma:Reviews1,ref:plasma:Reviews2,ref:plasma:Reviews3} The wide
interest to the problem is motivated by intriguing chance to create a new state
of matter, the quark-gluon plasma (QGP), in extraordinary hot and dense
environment, which is expected to be formed in relativistic collisions of
heavy nuclei.

This new state of matter is characterized by absence of the color confinement.
Indeed, in the QGP the quarks and the gluons are freely propagating particles while in
the hadron phase these colored degrees of freedom are
tightly bound into hadrons and glueballs. Another interesting feature of
hot QGP is the restoration of the chiral symmetry as
the chiral condensate melts down at sufficiently high temperatures.

Clearly, both deconfining and chiral transitions must happen somewhere between
hadronic and QGP phases, and they are not necessarily related to each other.
Recently, lattice QCD simulations revealed that at zero baryon density
the chiral restoration transition and the deconfinement transition may presumably occur separately:
in the continuum limit the restoration of the chiral symmetry may happen at a lower temperature
compared to the temperature at which the color deconfinement takes place.\cite{ref:Fodor}
In realistic QCD, however, the chiral and deconfining transitions
are unlikely to be real {\it phase} transitions. These transitions are rather smooth
crossovers characterized by analytical behavior of thermodynamic quantities across
transition(s), and the meaning of the transition temperature(s)
is somewhat blurry. However, it seems natural to identify the temperature of the chiral
transition with a temperature at which the susceptibility of a light quark condensate
takes it maximum value. The deconfinement temperature can similarly be located with the
help of the Polyakov loop susceptibility.\cite{ref:Fodor} The choice of these quantities
is motivated by the fact that the expectation values of the quark condensate and the
Polyakov loop become exact order parameters in the chiral limit and in the limit of
the infinitely heavy quarks, respectively. See Ref.~\refcite{ref:Fodor} for further review and
references.

If true, the presence of the two distinct transitions in QCD -- associated with the chiral
symmetry restoration and with the color deconfinement -- indicates that the mechanisms of the
chiral symmetry breaking and the mechanism of the quark confinement may differ from each other.
The confinement of quarks is usually associated
with condensation of certain magnetic gluon configurations (called ``Abelian monopoles'').
In this mechanism of confinement -- which is often called the dual
superconductor scenario\footnote{This scenario
is related to another approach based on percolation of
vortexlike magnetic struc\-tu\-res.\cite{ref:Greensite} The Abelian monopoles and the
center vortices are geometrically linked to each other.~\cite{ref:Greensite,ref:relation1,ref:relation2}}
-- the quark confinement is guaranteed by emergence of a confining flux tube
stretched between a quark and an antiquark.\cite{ref:tHooft,ref:Mandelstam,ref:Nambu:conf} The
tube appears due to a dual analogue of the Meissner effect: the (chromo)electric field
of the quarks is squeezed into the flux tube as a result of the condensation of
the Abelian monopoles (for a review see Refs.~\refcite{ref:Review:Monopoles:1,ref:Review:Monopoles:2}).

Below we discuss a chiral counterpart of the Abelian monopole, which we call
a chiral monopole. We suggest that the dynamics of the chiral monopoles
\begin{enumerate}
\item[(i)] causes the restoration of the chiral symmetry in the QGP phase;
\item[(ii)] is not related, at least directly, to the color (de)confinement.
\end{enumerate}

We start from a brief description of the 't~Hooft--Polyakov (HP) monopole in the Georgi-Glashow (GG)
model\cite{ref:tHooft:mon,ref:Polyakov:mon} (Section 2). In Section 3
we describe the chiral (or, ``quark'') monopole which is a QCD analogue of the HP
monopole.\cite{ref:Chernodub} Finally, in Section 4 we discuss the mentioned link between
the dynamics of the chiral monopoles and the chiral symmetry restoration at high temperature.

\section{'t~Hooft--Polyakov monopoles in Georgi-Glashow model}

As an illustrative example let us consider the GG model,
\beqn
\cL_{\mathrm{GG}} = - \frac{1}{4} {\vec G}_{\mu\nu} \cdot {\vec G}^{\mu\nu}
+ \frac{1}{2}  D_\mu \vec \Phi \cdot D^\mu \vec \Phi
- \frac{\lambda}{4} [\vec \Phi^2 - \eta^2]^2\,.
\label{eq:GG}
\eeqn
This model describes dynamics of the $SO(3)$ gauge field $A^a_\mu$ coupled to the triplet
Higgs field $\Phi^a$, $a=1,2,3$ via the adjoint covariant derivative
\beqn
{(D_\mu)}^{ab} = \delta^{ab}\, \partial_\mu + g \, \epsilon^{abc} A^c_\mu\,,
\label{eq:D:adjoint}
\eeqn
where $g$ is the gauge coupling and
$\vec G_{\mu\nu} = \partial_{[\mu,} \vec A_{\nu]} + g \vec A_\mu \times \vec A_\nu$
is strength tensor of the gauge field. The scalar coupling $\lambda$ controls
self-interaction of the scalar field and the condensate of the scalar field
is $|\langle \vec \Phi \rangle| = \eta$.

The HP monopole is described by the ansatz\cite{ref:tHooft:mon,ref:Polyakov:mon}:
\beqn
\Phi^a =\frac{r^a}{g\, r^2} H(\eta g\, r)\,, \qquad
A^a_i = \epsilon_{aij} \frac{r^j}{g\, r^2} [1 - K(\eta g\, r)]\,, \qquad
A^a_0 = 0\,,
\label{eq:Phi}
\eeqn
where $K$ and $H$ are two profile functions which can be found by solving
classical equations of motion of the model~\eq{eq:GG}.

For a static monopole, the field $\vec \Phi$ has a hedgehoglike structure with respect
to the spatial vector, $\vec \Phi \propto \vec r$. Since $\vec\Phi$ is a single-valued field
then the scalar condensate in the geometrical center of the monopole should vanish, $|\vec\Phi(0)| = 0$.
In other words, the core of the HP monopole destroys the Higgs condensate.

In quantum ensembles the behavior of the gauge field and the scalar field around an
HP monopole is obviously different from the ansatz~\eq{eq:Phi}. Moreover, the HP
monopoles are in general non-static. Thus, in order to determine the HP monopoles
in nonclassical field configurations one needs a gauge-- and Lorentz--invariant criterium.
To this end we need to know the 't~Hooft tensor\cite{ref:tHooft:mon}
\beqn
\cF_{\mu\nu}(n,A) = \vec G_{\mu\nu}(A) \cdot \vec n -
\frac{1}{g} {\vec n} \cdot D_\mu \vec n \times D_\nu \vec n\,,
\qquad
\vec n = \frac{\vec \Phi}{|\vec \Phi|}\,,
\label{eq:thooft:tensor:def}
\eeqn
where the unit vector $\vec n$ points towards the color direction of the triplet scalar field.
The 't~Hooft tensor~\eq{eq:thooft:tensor:def} is the gauge-invariant field strength tensor
corresponding to the composite Abelian gauge field $\cA_\mu = \vec A_\mu \cdot \vec n$.

The monopole current can now be determined by a Maxwell equation
\beqn
k_\nu = \frac{g}{4 \pi} \partial_\mu \tilde \cF_{\mu\nu}\,, \qquad
\tilde \cF_{\mu\nu} = \frac{1}{2} \epsilon_{\mu\nu\alpha\beta} \cF_{\alpha\beta}\,.
\label{eq:k:def}
\eeqn
By definition, the current of the HP monopole,
\beqn
k_\nu = \int_{\cC} d\tau \frac{\partial X^{\cC}_\nu(\tau)}{\partial \tau}\,
\delta^{(4)}(x - X^{\cC}(\tau))\,,
\eeqn
has a delta--like singularity at the closed monopole worldline $\cC$,
parameterized by the four-vector $X^{\cC}(\tau)$. Equations~\eq{eq:thooft:tensor:def}
and \eq{eq:k:def} guarantee that the monopoles are quantized and that the monopole
charge is conserved. If one applies Eq.~\eq{eq:k:def} to the HP ansatz~\eq{eq:Phi}
then one gets the static current $k_\mu = \delta_{\mu0} \, \delta^{(3)} (\vec r)$.

The HP monopoles have interesting dynamical and kinematical properties.
These objects were first formulated as the classical solutions in the Higgs (broken)
phase of the GG model.\cite{ref:tHooft:mon,ref:Polyakov:mon} In this phase the non-Abelian
symmetry of the model is broken down by the Higgs condensate to its Abelian subgroup, $SO(3) \to SO(2)$.
The HP monopoles are rare objects in the Higgs phase (this property of the model is consistent with
the fact that the Higgs condensate is destroyed in the cores of the monopoles).
However, the GG model can also reside in a symmetric (confining) phase in which the
symmetry is unbroken, the Higgs condensate is absent and the density of the HP
monopoles is high.

The physical picture of the phase transition from the broken (Higgs)
phase into the symmetric phase can be interpreted in terms of
the HP monopoles: as we move along a certain path in the coupling space
starting from the broken phase towards the symmetric phase, the
density of the monopoles increases\footnote{For simplicity we discuss
the model at zero temperature in order to avoid inessential details related
to appearance of the another symmetric phase (``Coulomb phase'') which is deconfining.
This Coulomb phase is also characterized by enhanced -- with respect to the broken phase --
density of the HP monopoles, which are, however, not condensed.}.
The Higgs condensate melts inside the monopole cores, and, as a result, the bulk
expectation value of the Higgs condensate lowers with the increase of the monopole density.

At some point of our path the Higgs condensate disappears. This point
corresponds to a transition separating the broken phase
from the symmetric phase (i.e., at this point our path touches the boundary between the phases).
The transition point corresponds to the critical density of the HP monopoles at which
\begin{enumerate}
\item[(i)] the HP monopoles start to condense in a given environment;
\item[(ii)] the color symmetry gets restored.
\end{enumerate}
The symmetric phase is filled by the monopole condensate, which is
absent in the Higgs phase. As we continue to move in the symmetric phase
outwards the broken phase, the Higgs condensate stays zero while the
monopole condensate strengthens.

We would like to apply the described monopole-mediated scenario
to QCD in order to describe the chiral symmetry restoration. However, a
{\it direct} application of the HP construction to QCD seems to be
impossible\footnote{In (pure) non-Abelian gauge theories one can still construct an
effective Higgs field via an Abelian gauge fixing and then identify
Abelian monopoles relevant to the confinement of color.\cite{ref:Review:Monopoles:1,ref:Review:Monopoles:2}}
since there are no scalar fields in QCD, and, moreover, there is no evidence that
the color symmetry is broken at low baryon density.
Nevertheless, monopolelike defects of a HP type
can be identified in a non-Abelian gauge theory with (generally, dynamical)
quarks, and, as we argue below, the dynamics of these monopoles may
indeed be related to the chiral symmetry breaking/restoration.

\section{Chiral monopoles in QCD}
\label{sec:construction}

For simplicity let us consider the $SU(2)$ gauge theory with one
quark field~$\psi$ which transforms in the fundamental representation of the gauge
group. Generalizations to a multiflavor theory\cite{ref:Chernodub} and to the
case of the $SU(3)$ group\cite{ref:Chernodub:Morozov} are straightforward.

The key idea is to use the quark field $\psi$ in order to construct
a composite scalar field $\vec \xi$ transforming in the adjoint representation
of the gauge group. The field $\vec \xi$ should then play a role
which is similar to the role the scalar field $\vec \Phi$ in the
GG model. This composite field can be constructed in various
ways. A simplest choice is given by the quark-antiquark bilinears,
which can formally be written as:
\beqn
\vec \xi_{\Gamma} = \bar\psi(x) \Gamma \vec \tau \psi(x)\,,
\qquad \Gamma = \bbbone\,, i \gamma_5\,.
\label{eq:xi}
\eeqn
Here $\vec \tau=(\tau_1,\tau_2,\tau_3)$ are the Pauli matrices
acting in the color space and $\gamma_\mu$, $\gamma_5$ is the
standard set of the spinor $\gamma$--matrices. The
real-valued fields $\vec \xi_S$ and $\vec \xi_A$ transform
with respect to the rotations/reflections of the coordinate space
as the scalar field and the pseudoscalar (axial) field,
respectively. The subscripts $S$ and $A$ correspond, respectively, to
the scalar, $\Gamma = \bbbone$, and axial, $\Gamma = i \gamma_5$, operators.

In order to make the definition~\eq{eq:xi} meaningful one should consider, for example,
the fermion field~$\psi$ as a $c$-valued function. It is convenient
to choose the field~$\psi$ to be an eigenmode $\psi_\lambda$ of the Dirac operator
$\cD \equiv \cD[A]$, which corresponds to a given background configuration of
the gauge field $A_\mu$:
\beqn
\cD[A] \psi_\lambda(x) = \lambda \psi_\lambda(x)\,, \qquad
\cD[A] = \gamma_\mu (\partial_\mu + i \frac{1}{2} \tau^a A^a_\mu) + m\,,
\label{eq:Dirac:equation}
\eeqn
The Dirac eigenmodes are labeled by the eigenvalues $\lambda$ of the Dirac operator, so that
one can identify infinite number of the effective adjoint composite fields corresponding to a
given background gauge field $A$,
\beqn
\vec \xi_{\Gamma,\lambda} = \bar\psi_\lambda(x; A) \Gamma \vec \tau \psi_\lambda(x; A)\,.
\label{eq:xi:lambda}
\eeqn
The eigenvalue index $\lambda$ of the composite fields~\eq{eq:xi:lambda}
corresponds to a virtual energy scale which is ``resolved'' by the
composite field $\vec \xi_{\Gamma,\lambda}$.

One can also define the composite scalar field $\vec \xi$ as an average of a
quark-antiquark bilinear over all possible fermion fields in the background of the
given gauge field $\vec A_\mu$. This composite field can be
represented a (local) quark condensate
\beqn
\vec \xi^{(\psi)}_{\Gamma}[A] = \left\langle \bar\psi(x) \Gamma \vec \tau \psi(x) \right\rangle_A
\equiv
\sum_\lambda \frac{\bar\psi_\lambda(x; A)\Gamma \vec \tau \psi_\lambda(x; A)}{\lambda - i m}\,.
\label{eq:xi:average}
\eeqn
To our mind this equation represents the most suitable choice of the
composite adjoint scalar field. First of all, the composite field~\eq{eq:xi:average}
is written is a simple and natural way. There is no predistinguished energy
scale $\lambda$ labeling this field. Finally, the nonperturbative (infrared) eigenmodes
enter Eq.~\eq{eq:xi:average} with a higher weight compared to the perturbative (ultraviolet)
modes. For shortness, we use below the notation $\vec \xi_{\Gamma}$
for both definitions \eq{eq:xi:lambda} and \eq{eq:xi:average}.

Under the axial transformations ($\alpha$ is the global parameter of the axial rotation),
\beqn
U_A(1): \qquad \psi \to e^{i \alpha \gamma_5} \psi\,, \quad
\bar\psi \to \bar\psi e^{i \alpha \gamma_5}\,,
\label{eq:axial:U1}
\eeqn
the color vectors $\vec \xi_S$ and $\vec \xi_A$ transform via each other:
\beqn
\left(\begin{array}{c} \vec \xi_S \\ \vec \xi_A \end{array} \right) \to
{\left(\begin{array}{c} \vec \xi_S \\ \vec \xi_A \end{array} \right)}'
=
\left(\begin{array}{cc}
\cos 2 \alpha & \sin 2 \alpha \\ - \sin 2 \alpha & \cos 2 \alpha
\end{array}
\right)
\left(\begin{array}{c} \vec \xi_S \\ \vec \xi_A \end{array} \right)\,.
\label{eq:axial:SO2}
\eeqn

Using two adjoint fields $\xi_\Gamma$ we can define three unit color vectors
\beqn
\vec n_S = \frac{\vec \xi_S}{|\vec \xi_S|}\,, \quad
\vec n_A = \frac{\vec \xi_A}{|\vec \xi_A|}\,, \quad
\vec n_I = \frac{\vec \xi_S \times \vec \xi_A}{|\vec \xi_S \times \vec \xi_A|}\,.
\label{eq:ns}
\eeqn
Here $(\vec u, \vec v)$ and $[\vec u \times \vec v]^a = \epsilon^{abc} u^b v^c$
are, respectively, the scalar and the vector products in the color space
of the vector $\vec u$ and $\vec v$, and
$|\vec u| = (\vec u, \vec u)^{1/2}$ is the norm of the color vector $\vec u$.
The vector $\vec n_I$ is invariant under the axial
transformations~(\ref{eq:axial:SO2}) because it is a (normalized) vector
product of the scalar and axial vectors.

Our definition of the composite adjoint fields $\vec \xi_\Gamma$ inherently
``locks'' the axial rotations with the gauge rotations since
Eq.~\eq{eq:axial:SO2} may also be regarded as a global
rotation in a color (gauge group) space by the angle $2 \alpha$
around the color direction~$\vec n_I$.

Now we interpret the unit vectors~\eq{eq:ns} as directions of
certain composite adjoint Higgs fields. We have three sets
of the fields $\{\vec n_\Gamma, \vec A_\mu\}$ with $\Gamma=S,A,I$,
which can be used to construct three gauge invariant 't~Hooft
tensors~\eq{eq:thooft:tensor:def} as it was already done
in Section~2 for the case of the GG model:
\beqn
\cF^\Gamma_{\mu\nu}(n_\Gamma,A) = F^a_{\mu\nu}(A)\, n_\Gamma^a -
\frac{1}{g} \epsilon^{abc} n_\Gamma^a {(D^{\mathrm{ad}}_\mu
n_\Gamma)}^b {(D^{\mathrm{ad}}_\nu n_\Gamma)}^c\,,
\qquad
\Gamma = S\,,A\,,I\,.
\label{eq:thooft:tensor}
\eeqn

The 't~Hooft tensor~\eq{eq:thooft:tensor} is the gauge-invariant field strength tensor for the
diagonal (with respect to the color direction $\vec n_\Gamma$) component of the gauge
field,
\beqn
\cA^\Gamma_\mu = A^a_\mu n^a_\Gamma\,, \qquad \Gamma = S\,,A\,,I\,.
\label{eq:A:Gamma}
\eeqn

We now come close to the definition of the chiral monopole(s).
The current of the chiral monopole of the $\Gamma$-th type is
\beqn
k^\Gamma_\nu = \frac{g}{4 \pi} \partial_\mu \tilde \cF^\Gamma_{\mu\nu}\,,
\label{eq:k:Nambu}
\eeqn
where we used Eq.~\eq{eq:k:def} derived in the GG model. The currents~\eq{eq:k:Nambu}
have delta-like singularities at the corresponding worldlines.
The monopole charges of the chiral monopoles -- defined according to Eq.~\eq{eq:k:Nambu} --
are quantized and conserved.
We would like to stress that the chiral monopoles in QCD are explicitly gauge invariant.

According to Eq.~\eq{eq:A:Gamma}
the chiral monopoles of the $\Gamma$-th type carry the magnetic charges with
respect to the ``scalar'' ($\Gamma=S$), ``axial'' ($\Gamma=A$) and ``chirally invariant'' ($\Gamma=I$)
components of the gauge field $\vec A_\mu$. In the corresponding Unitary gauges,
$n^a_\Gamma = \delta^{a3}$, the quark monopoles correspond to
monopoles ``embedded'' into the diagonal
component~\eq{eq:A:Gamma}. In the gauges, where the diagonal
component $\cA^\Gamma_\mu$ is regular, such monopoles are
hedgehogs in the composite quark-antiquark fields
(colored quark condensates). Each monopole
is characterized by the typical hedgehoglike
behavior ($\vec n_\Gamma \sim \vec r$ for static monopoles)
in the local vicinity of the monopole core. The very existence
of these monopoles in QCD is not a dynamical fact but rather
a simple kinematical consequence of the existence of the
adjoint real-valued fields defined via Eqs.~(\ref{eq:xi}) and (\ref{eq:ns}).

Thus, the chiral monopole is gauge-invariant hedgehoglike structure in the ``colored quark condensate''
\eq{eq:xi:lambda} or \eq{eq:xi:average}. The composite nature
of the condensate does not undermine the existence of the chiral monopoles. For example, a very
similar structure, called the (embedded) Nambu monopole, is a well known field
defect in the Standard Electroweak theory.\cite{ref:Nambu} In the Electroweak theory the role of the adjoint
composite field~\eq{eq:xi} is played by the scalar triplet
$\vec \xi_{\mathrm{EW}} = \Phi^\dagger \vec\tau \Phi$ field,
where $\Phi$ is the two-component Higgs field. A similar type
of defects exists also in a superfluid Helium as well as in
certain types of liquid crystals.\cite{ref:Volovik}

\section{Chiral monopoles and chiral symmetry restoration}

Having in mind the analogy between the chiral monopole in QCD
and the HP monopole in the GG model, one can suggest that
the properties of the chiral monopoles are related
to the restoration of the chiral symmetry in the high-temperature phase of QCD.\cite{ref:Chernodub}
The chain of considerations is as follows.

Firstly, the cores of the chiral monopoles should contain a chirally symmetric
vacuum (this statement is intuitively clear
because of the hedgehoglike structure of the local condensates).\cite{ref:Chernodub}
Secondly, it was found numerically that the density of
the chiral monopoles increases with temperature.\cite{ref:Chernodub:Morozov}
One can interpret this observation as a destruction of the bulk expectation
value of the chiral condensate by the cores of the chiral monopoles:
as density of the chiral monopole increases, the vacuum of QCD gradually
turns from the chirally broken phase to the chirally symmetric phase. Thirdly, another
numerical argument in favor of our conclusion can be found with the help of
the mode-by-mode analysis: the density of the chiral
monopoles is anti-correlated with the density of the Dirac eigenmodes (the lower
density of the Dirac eigenmodes the higher monopole density).\cite{ref:Chernodub:Morozov}

Most probably the chiral monopoles are not related to the confining properties
of QCD. Indeed, the confinement of color needs a certain amount of disorder in
the gauge fields. The disorder is usually reflected in existence of a kind of a condensate made of percolating
defect trajectories. In other words, the defects, which presumably cause the
confinement, are to be propagating (proliferating) for infinitely long
distances in the confinement phase. The condensate of the defects must disappear in
the deconfinement phase. These properties were indeed observed in $SU(2)$ Yang--Mills theory
both for the Abelian monopoles\cite{ref:Review:Monopoles:1,ref:Review:Monopoles:2} and
for the center vortices\cite{ref:Greensite}, both of which are the most probable candidates for
the confining gluonic configurations. However, the chiral monopoles are suggested
to be percolating in the chirally symmetric QGP phase which is, however, not confining.\cite{ref:Chernodub}
On the contrary, in the QGP phase the Abelian monopoles may form a gaseous or a liquid state,
rather than a condensate.\cite{ref:Chernodub:Zakharov} Thus, the properties of the confining
(Abelian) monopoles and the chiral monopoles are rather different.

The chiral monopole is an object made of both quark
and gluon fields. Therefore, one can expect that the presence of the chiral monopole affects not only
the quark condensate in the local vicinity of the monopole, but the chiral monopole
may also have its imprint on the gluonic
fields (in a complete analogy with the Abelian monopoles,
see Refs.~\refcite{ref:Bakker,ref:Anatomy} for details).
Indeed, it was found numerically in Ref.~\refcite{ref:Chernodub:Morozov}
that the chiral monopoles possess gluonic cores: on average, the chromomagnetic energy
near the monopole trajectories is higher compared to the
chromomagnetic energy far from the monopole cores. Features of the gluonic core
are consistent with the asymptotic freedom. The clear
gluonic structure of the chiral monopole in QCD makes this monopole very similar
to the HP monopole in the GG model.

Summarizing, we suggested a scenario of the chiral symmetry restoration
in the QGP phase. We argued that the increase of the density
of the chiral monopoles with temperature leads to
the gradual suppression of the chiral condensate in this phase. At the same time, the
chiral monopoles seem to be unrelated to the confinement of color, so that
the chiral symmetry restoration and the deconfinement transition may occur at different
temperatures.

\section*{Acknowledgments}
Author is thankful to organizers of the conference (ITEP, Moscow)
and the workshop (RCNP, Osaka) for kind hospitality and fruitful atmosphere.
The work is supported by Federal Program of the Russian Ministry of Industry,
Science and Technology No. 40.052.1.1.1112, by the grants RFBR 07-02-08671-z,
RFBR-DFG 06-02-04010, by a STINT grant IG2004-2 025 and by a CNRS grant.

\end{document}